\documentclass[aps,twocolumn]{revtex4}
\usepackage{amssymb}
\usepackage{epsfig}
\usepackage{amsmath}
\usepackage{graphicx}

\begin{document}

\title{Enhancing robustness and immunization in geographical networks}
\author{Liang Huang,$^{1,3}$\footnote[1]{Present address: Department of Electrical Engineering,
Arizona State University, Tempe, Arizona 85287, USA.} Kongqing
Yang$^{2,3}$ and Lei Yang$^{1,3,4}$\footnote[2]{Corresponding
author.}} \affiliation{$^{1}$Institute of Modern Physics, Chinese
Academy of Science, Lanzhou 730000, China}
\affiliation{$^{2}$Institute of Applied Physics, Jimei University,
Xiamen 361021, China} \affiliation{$^{3}$Department of Physics,
Lanzhou University, Lanzhou 730000, China} \affiliation{$^{4}$The
Beijing-Hong Kong-Singapore Joint Centre for Nonlinear and Complex
Systems (Hong Kong), Hong Kong Baptist University, Hong Kong,
China}

\begin{abstract}
We find that different geographical structures of networks lead to
varied percolation thresholds, although these networks may have
similar abstract topological structures. Thus, the strategies for
enhancing robustness and immunization of a geographical network
are proposed. Using the generating function formalism, we obtain
the explicit form of the percolation threshold $q_{c}$ for
networks containing arbitrary order cycles. For 3-cycles, the
dependence of $q_c$ on the clustering coefficients is ascertained.
The analysis substantiates the validity of the strategies with an
analytical evidence.
\end{abstract}

\date{\today }
\maketitle

Complex networks (see reviews \cite{CN-review}) provide powerful
tools to investigate complex systems in nature and society. The
properties of complex systems are affected by the geographical
distribution of the components. For example, routers of the
Internet \cite{internet} and transport networks \cite{transport}
lay on the two-dimensional surface of the globe; world-wide
airport network is confined by the geography \cite{airport};
neuronal networks in brains \cite{neuron} occupy three-dimensional
space. Thus it is helpful to study the geographical complex
networks \cite{BBPV:2004,lesf,snsf,GCN-other,GCN-yang,sandpile}.

From the abstract geometrical point of view, an abstract set can
describe a general system. When it is equipped with some
geometrical structures, the set can further describe a specified
system. So, an abstract topological network is an abstract set
that consists of nodes and links. A metric can be added to the
abstract topological network, the metric could be arbitrary, not
necessarily the Euclidean metric \cite{BBPV:2004}. Embedding a
particular abstract topological network, into a suitable metric
space provides a method to add a metric.

In this paper, the problem of percolation thresholds in
geographical networks is studied. Three types of geographical
networks are investigated: the normal model \cite{lesf,snsf}, the
hollow model and the concentrated model. By extensive numerical
simulations, we found that the percolation threshold $q_{c}$ (the
point that a spanning connected cluster emerges) for these models
satisfies: $q_{c}$(concentrated)$>$$q_{c}$(normal)$>$$
q_{c}$(hollow). Based on these results, we suggest a strategy
(hollowing) for enhancing robustness and a strategy
(concentrating) for immunization of geographical networks. The
different geographical networks have different distribution of
cycles. The geographical dependence of the percolation threshold
is investigated by the generating function process in abstract
networks containing cycles.

Based on the lattice embedded model \cite{lesf}, the networks are
generated as follows: each node in an $L\times L$ lattice with
periodic boundary conditions is assigned a degree quota $k$, drawn
from the prescribed degree distribution. Here, scale-free
($P(k)\sim k^{-\lambda },m\leqslant k$) and exponential ($P(k)\sim
e^{-k/k_{0}},m\leqslant k$) degree distributions are considered,
where $m$ is the minimum degree a node can have. (a) The normal
case [lattice embedded scale-free (LESF) or lattice embedded
exponential (LEE)]: A node $i$ connects to its closest neighbors
until its degree $ k_{i}$ is realized, or up to a cutoff distance
$A\sqrt{k_{i}}$, where $A$ is large enough to ensure that almost
all the degree quotas can be fulfilled. (b) The hollow case
[hollow LESF (HLESF) or hollow LEE (HLEE)]: similar to the normal
case, except that a node $i$ has probability $p$ to be forbidden
to connect its first $n$ nearest neighbors. (c) The concentrated
case: $A$ is set smaller than that in the normal case. The process
is repeated throughout all the nodes in the lattice. For the
hollow case, when the network degenerates to a lattice and $n$ is
small, it becomes similar to the tunneling effect on Euclidean
lattices \cite{GS1995}. In the following simulations, we choose
$A=7$, $n=8 $, and $p=1$; network size $N=10^{6}$, minimum degree
$m$ is $4$ for scale-free networks and $1$ for exponential
networks, and all the data are averaged over $1000$ ensembles,
unless otherwise specified.

\begin{figure}[th]
\centering \epsfig{file=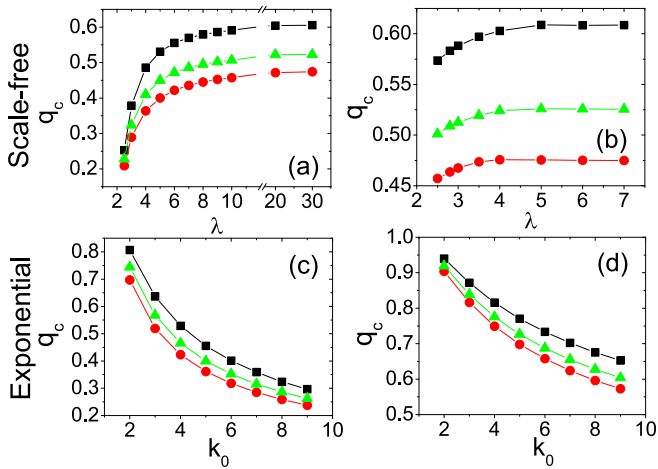, width=\linewidth}
\caption{(Color online) Percolation thresholds of the networks for
both random failures (left panels) and intentional attacks (right
panels). In each subgraph squares represent for normal lattice
embedded networks, triangles for hollow lattice embedded networks
with $p=0.5$ and circles for $p=1$.} \label{qcle}
\end{figure}

\begin{figure}[th]
\centering \epsfig{file=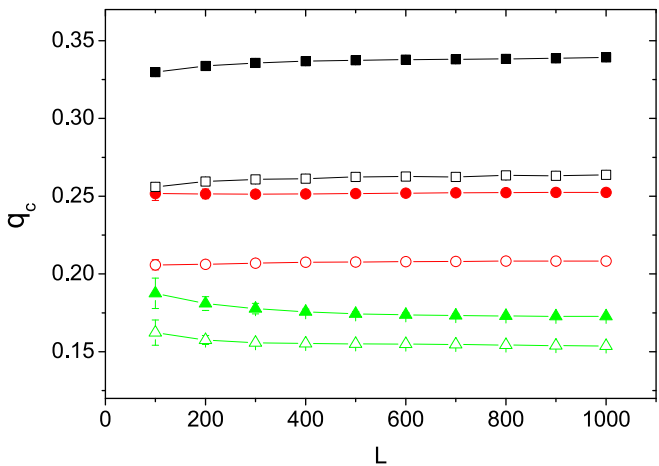, width=\linewidth}
\caption{(Color online) Random percolation threshold $q_{c}$ vs
network side length $L$ for the LESF model (filled symbols) and
the HLESF model (empty symbols) for different $\protect\lambda $:
Squares for $\protect\lambda =2.8$ , circles for $\protect\lambda
=2.5$, and triangles for $\protect\lambda =2.3 $. Each data is the
result of averaging $10^{4}$ network realizations.} \label{qc23}
\end{figure}

The algorithms of Newman and Ziff \cite{algorithm} is performed to
calculate the threshold $q_{c}$, which is defined as the point
where the differential of the size of the largest cluster as a
function of occupying probability $q$ maximizes. The definition is
equivalent to the usual definition by the emergence of a spanning
cluster for large network size limit \cite{threshold}. Figure
\ref{qcle} shows a clear drop of the percolation threshold $q_{c}$
in the hollow networks than in the normal networks. So the
robustness can be enhanced by the model. The size effect
(scale-free degree distribution) is demonstrated in Fig.
\ref{qc23}. The LESF and even the HLESF networks also have
non-zero percolation thresholds for $\lambda \in (2,3)$, the
results are consisitent with Ref. \cite{threshold}. Again, the
drop in $q_{c}$ for hollow networks is apparent.

\begin{figure}[th]
\centering \epsfig{file=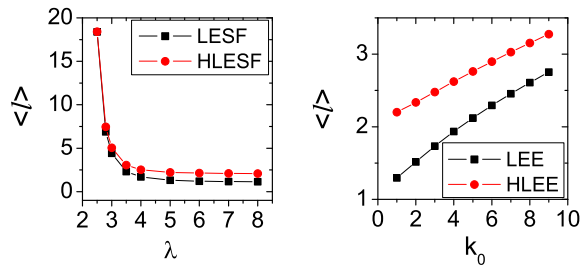, width=\linewidth}
\caption{(Color online) The average spatial length of connections
of the normal networks (squares) and hollow networks (circles) for
scale-free degree distribution (a) and exponential degree
distribution (b).} \label{len}
\end{figure}

The amendment in hollow networks is small in the physical space.
As Fig. \ref{len} shows, the average spatial length $\left\langle
l\right\rangle $ of the edges for the hollow networks dose not
increase much compared with the normal networks, for small
$\lambda $ or large $k_{0}$. As the average degree $\left\langle
k\right\rangle $ becomes larger, the difference between
$\left\langle l\right\rangle $\ decreases. In the limit case, the
difference goes to $0$. While the cost of constructing hollow
networks still remains low, they are more robust than the normal
models. Under the same conditions, the hollow networks have
significant lower percolation thresholds.

\begin{figure}[th]
\centering \epsfig{file=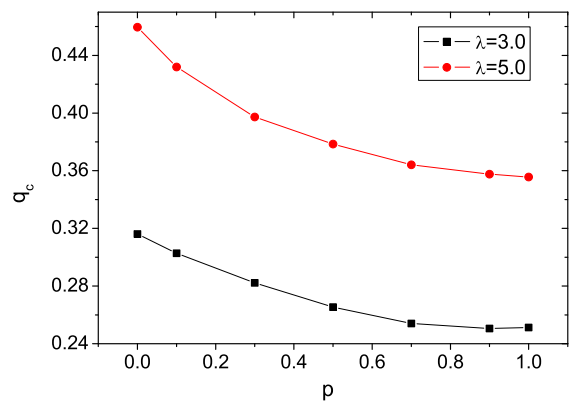, width=\linewidth}
\caption{(Color online) Random percolation thresholds vs the
rearrange probability for LESF networks, $n=8$.} \label{hs}
\end{figure}

Based on the above observations, we propose a hollowing strategy
to enhance the robustness of geographical networks. For each node
in a geographical networks, we introduce a probability $p$ to cut
down the edges that linked to its first $n$ nearest neighbors,
then to reconnect further nodes in the geographical distance. In
this case, the degree distribution $P(k)$ deviates a little for
small $k$, i.e., around about $10$. This only causes a variation
in percolation thresholds of a much smaller magnitude. Figure
\ref{hs} demonstrates the efficiency of the hollowing strategy for
LESF networks. The percolation threshold drops about $0.1$ with
$n=8$, namely, it needs 10 percent less nodes of the network to
maintain a spanning cluster. This can have significant effect to
prevent the network from breakdown when the network undergoes a
serious crisis and is losing its global function.

\begin{figure}[th]
\centering \epsfig{file=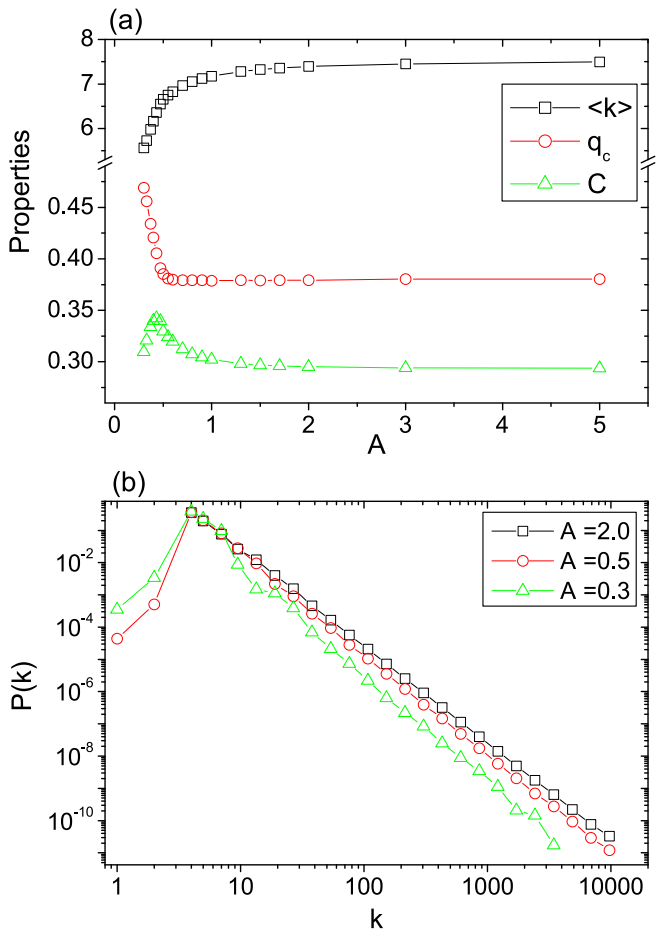, width=\linewidth}
\caption{(Color online) Properties of concentrated networks. (a):
Squares are the average degrees, circles are the random
percolation thresholds, and triangles are the clustering
coefficients. (b): The actual degree distribution for three
typical $A$ values. The data are log-binned. The assigned degree
distribution is scale-free, with $\protect\lambda =3.0$.}
\label{con}
\end{figure}

For concentrated cases, we study the concentrated scale-free
networks since the problem mainly concerns with immunization
strategies \cite{YCLai}. The long-range links (for large $A$) in
the LESF model are different from those in the small-world model
\cite{sw}. In the LESF model, a node $i$ first tries to connect to
its nearest neighbors; if they have already fulfilled their degree
quotas, this node has to try further nodes until up to the cutoff
distance $A\sqrt{k_i}$. The long-range links in the LESF model are
still somewhat local. As far as $A$ is large enough to ensure that
almost all the nodes' degree quotas are fulfilled, the network
properties depend little on $A$, as shown in Fig. \ref{con}(a).
However, as $A$ decreases, more and more long-range links are
prohibited \cite{lesf}. The network becomes more localized, thus
the clustering coefficient increases while the average degree
decreases a little. As $A$ is approximately larger than $0.5$,
most of the nodes' degree quotas are still fulfilled except the
nodes with large degree quotas. Thus the network structure remains
mainly unchanged, and the percolation thresholds remains almost
the same. As $A$ decreases further ($A < 0.5 $), a large portion
of the degree quotas of nodes cannot be fulfilled. Figure
\ref{con}(b) shows a serious degree cutoff ($A = 0.3$), where the
steps in the degree distribution come from the symmetry of the 2D
lattice. Thus the network structure becomes seriously
deteriorated: the average degree decreases sharply, the clustering
coefficient drops; the percolation threshold increases rapidly.
These effects indicate that the concentrated networks, are much
easier to be immunized. This is indeed the case that, during an
epidemic, most people are staying home to seclude from the
infection.

To better understand the numerical results, we employ the
generating function method to determine the percolation threshold
for networks with different clustering properties. Here the
clustering properties can be simply depicted by the clustering
coefficient, which counts for the triangles (3-cycles) in the
network, and it can also be represented by the number of
rectangles (4-cycles), and generally by the number of $L$-cycles.
In the following, a general relation of the percolation threshold
$q_c$ and the number of $L$-cycles for a random network with
arbitrary degree distribution is determined. As an example, the
dependence of $q_c$ on the clustering coefficient is obtained.

For uniform occupations (or random failures), the percolation
threshold of random tree-like networks is $q_{c}=\left\langle
k\right\rangle /\left\langle k(k-1)\right\rangle $
\cite{cohenqc,genth}. It could be obtained by the condition that
the average cluster size diverges, or equivalently, the average
size of clusters that reached by following an edge diverges. A
real network is usually clustered and contains certain amount of
cycles. If the number of cycles is small, (e.g. each node belongs
to at most one cycle) the generating function process can be
extended to cope with the random percolation problem.

For a uniform occupation probability $q$, the generating function
for the probability of the number of outgoing edges of a target
node reached by following a randomly chosen edge on a clustered
network remains the same as that of the random tree-like networks
\cite{genth,genth1}: $F_{1}(x)=\frac{q}{\left\langle
k\right\rangle }\sum kP(k)x^{k-1}$. However, if an outgoing edge
is not independent, i.e., it is terminated to a node that has
already been visited (having no contribution for reaching new
nodes), the node with degree $k$ will only have $k-2$ independent
outgoing edges. Thus it is convenient to define
\begin{equation*}
F_{1}^{(1)}(x)=\frac{q}{\left\langle k\right\rangle }\sum
kP(k)x^{k-2}=x^{-1}F_{1}(x),
\end{equation*}
as the generating function of the number of outgoing edges reaching new
nodes for such a target node.

Let $H_{1}(x)$\ be the generating function of the size
distribution of the cluster that reached by following an edge, by
following $2$ independent edges, the cluster size distribution is
generated by $H_{1}(x)^{2}$. But if the $2$ edges originate from a
common node and belong to an $L$-cycle, the generating function
should be $H_{1}^{(l_{1})}(x)H_{1}^{(l_{2})}(x)$, where
$l_{1}=[(L-1)/2]$, $l_{2}=L-1-l_{1}$. $\left[ g\right] $ is Gauss'
function which returns the integer part of $g$. $H_{1}^{(l)}(x)$
is the generating function for the size distribution of the
clusters that is reached by an edge and has one edge terminated
after $l$ steps, and satisfies an iterative relation
\begin{equation*}
H_{1}^{(l)}(x)=1-F_{1}(1)+xF_{1}^{(1)}(H_{1}(x))H_{1}^{(l-1)}(x),
\end{equation*}%
where the terminal condition $H_{1}^{(1)}(x)$ is
\begin{equation*}
H_{1}^{(1)}(x)=1-F_{1}(1)+xF_{1}^{(1)}(H_{1}(x)).
\end{equation*}%
Thus
\begin{eqnarray*}
H_{1}^{(l)}(x) &=&\left( 1-F_{1}(1)\right) \frac{1-\left(
xF_{1}^{(1)}(H_{1}(x))\right) ^{l}}{1-xF_{1}^{(1)}(H_{1}(x))} \\
&&+\left( xF_{1}^{(1)}(H_{1}(x))\right) ^{l}.
\end{eqnarray*}

In general, we may assume that a node $i$ with degree $k_{i}$ on
average belongs to $n_{L}(k)$ $L$-cycles, where $n_{L}(k)<1$ or
$\sim 1$. So $H_{1}(x)$\ satisfies the self-consistent equation
\begin{eqnarray*}
H_{1}(x) &=&1-F_{1}(1)+ \\
&&\frac{qx}{\left\langle k\right\rangle }\sum
kP(k)(H_{1}^{(l_{1})}(x)H_{1}^{(l_{2})}(x))^{n_{L}}H_{1}(x)^{k-1-2n_{L}}.
\end{eqnarray*}%
The average cluster size reached by an edge is%
\begin{eqnarray*}
\left\langle \widetilde{s}\right\rangle  &=&H_{1}^{\prime }(1) \\
&=&q+\frac{q}{\left\langle k\right\rangle }\sum
kP(k)\{n_{L}(k)(H_{1}^{(l_{1})\prime }(1)+H_{1}^{(l_{2})\prime }(1)) \\
&&+\left( k-1-2n_{L}(k)\right) H_{1}^{\prime }(1)\},
\end{eqnarray*}%
where
\begin{equation*}
H_{1}^{(m)\prime }(1)=\frac{1-q^{m}}{1-q}\left( xF_{1}^{(1)}(H_{1})\right)
^{\prime }(1)=\frac{1-q^{m}}{1-q}H_{1}^{(1)\prime }(1),
\end{equation*}%
and
\begin{equation*}
H_{1}^{(1)\prime }(1)=q+q\frac{\left\langle k(k-2)\right\rangle
}{\left\langle k\right\rangle }H_{1}^{\prime }(1).
\end{equation*}%
A simple substitution yields%
\begin{equation*}
\left\langle \widetilde{s}\right\rangle
=\frac{q+\frac{2-q^{l_{1}}-q^{l_{2}}}{1-q}\frac{q^{2}}{\left\langle
k\right\rangle }\left\langle kn_{L}\right\rangle
}{1-\frac{q}{\left\langle k\right\rangle }\{\left\langle
k(k-1)\right\rangle -2\left\langle
kn_{L}\right\rangle(1-\frac{2-q^{l_{1}}-q^{l_{2}}}{2(1-q)}\frac{q\left\langle
k(k-2)\right\rangle }{\left\langle k\right\rangle })\}}.
\end{equation*}%
Thus the percolation threshold $q_{c}$, given by the divergence of
$\left\langle \widetilde{s}\right\rangle $, is
\begin{equation}
q_{c}=\frac{\left\langle k\right\rangle }{\left\langle k(k-1)\right\rangle
-2\left\langle kn_{L}(k)\right\rangle \left( 1-\frac{%
2-q_{c}^{l_{1}}-q_{c}^{l_{2}}}{2(1-q_{c})}\frac{q_{c}\left\langle
k(k-2)\right\rangle }{\left\langle k\right\rangle }\right) }.  \label{qcl}
\end{equation}%
When $n_{L}=0$, this result degenerates to the known result of the
tree-like networks: $q_c=\left\langle k\right\rangle /\left\langle
k(k-1)\right\rangle $. In general $q_{c}<1$, as $L\rightarrow
\infty $, $q_{c}^{l_{1}}$ and $q_{c}^{l_{2}}$ limit to $0$, Eq.
\ref{qcl} reduces to a second order equation for $q_{c}$. The root
with physical meanings is $q_{c}=\left\langle k\right\rangle
/\left\langle k(k-1)\right\rangle $, which is just the percolation
threshold of the tree-like networks. This means that the cycles of
infinite length do not affect the percolation thresholds. The
numerical tests for several typical data values show that, for the
same number of cycles, $q_{c}$ is an increasing function of
$q_{c}^{l_{1}}+q_{c}^{l_{2}}$. The higher the cycle order $L$ is,
the less influence it will be. Thus the most influential cycles
are $3$-cycles, which could be expressed by the clustering
coefficients.

\begin{figure}[th]
\centering \epsfig{file=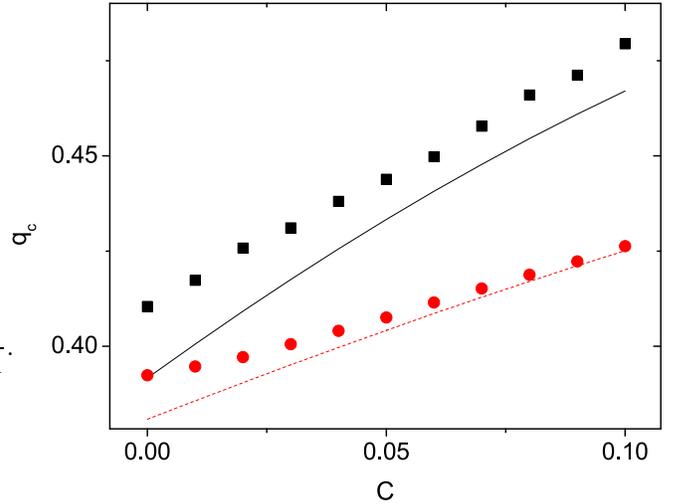, width=\linewidth}
\caption{(Color online) Percolation thresholds of slightly
clustered networks with truncated degree distribution $P(k)\sim
k^{-\protect\lambda }e^{-k/k_{0}}$, $k\geqslant m$. Squares:
$\protect\lambda =3$, $m=2$, $ k_{0}=10$; Circles:
$\protect\lambda =6$, $m=3$, $k_{0}=\infty $. The network size is
$10^{6}$, and each data is averaged over 1000 realizations. Solid
and dashed lines: theoretical result of Eq. (\protect\ref{qc}) for
$\protect \lambda =3$ and $\protect\lambda =6$ respectively.}
\label{qcn}
\end{figure}

For $L=3$, $l_{1}=l_{2}=1$. If the clustering coefficient
$C=\left\langle C(k)\right\rangle $ \cite{Ck} is small enough, we
may assume that two triangles could only have at most one common
node. A node with degree $k $ reached by an edge will belong to on
average $n_{3}(k)=\left(
C(k)(k-1)(k-2)+C(k)(k-1)\right) /2=C(k)(k-1)^{2}/2$ triangles. Equation \ref%
{qcl} will reduce to%
\begin{equation}
q_{c}=\frac{\left\langle k\right\rangle }{\left\langle k(k-1)\right\rangle
-(1-q_{c}\frac{\left\langle k(k-2)\right\rangle }{\left\langle
k\right\rangle })\left\langle C(k)k(k-1)^{2}\right\rangle }.  \label{qc}
\end{equation}%
The percolation threshold $q_{c}$ increases monotonically with $\left\langle
C(k)k(k-1)^{2}\right\rangle $. It is straightforward that when $C(k)\ $%
limits to $0$, $n_{3}\rightarrow 0$, $q_{c}$\ returns to $\left\langle
k\right\rangle /\left\langle k(k-1)\right\rangle $. On the other hand, if $%
\left\langle C(k)k(k-1)^{2}\right\rangle $ diverges, $q_{c}$\ maximizes to $%
\left\langle k\right\rangle /\left\langle k(k-2)\right\rangle$.
Figure \ref{qcn} shows the dependence of the percolation threshold
$q_c$ on the clustering coefficient $C$. Curves are from Eq.
\ref{qc} and the symbols are from numerical simulation. For
simulation, first a random network with prescribed degree
distributions is generated, then a rewiring process \cite{rewire}
is applied to achieve the required value of clustering
coefficient, while keep the degree distribution unchanged.
Percolation is performed using Newman and Ziff's algorithm
\cite{algorithm}.

\begin{figure}[th]
\centering \epsfig{file=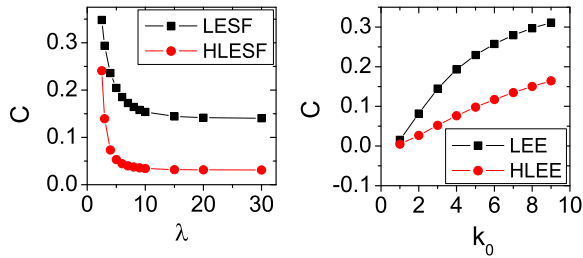, width=\linewidth}
\caption{(Color online) Clustering coefficients of the normal
(squares) and hollow (circles) networks with scale-free (left
panel) and exponential (right panel) degree distributions.}
\label{Cd}
\end{figure}

Although Eq. (\ref{qc}) holds only for the case of small
clustering coefficient $C$, the analysis above indicates that for
large $C$, or if the network has higher order cycles, the fraction
of edges that inter-connect existed nodes in a local cluster will
further increase. Thus the number of efficient edges connecting to
``new'' nodes (in comparison to the nodes within the local
cluster) may be even smaller, which results in a higher
percolation threshold. Thus when a network is more clustered, or
has more cycles---not only of order $3$---it will be less robust.

\begin{figure}[th]
\centering \epsfig{file=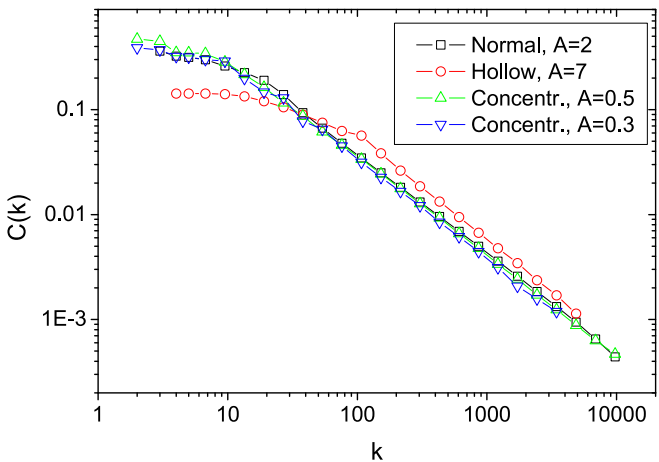, width=\linewidth}
\caption{(Color online) Clustering coefficient spectrum for the
normal model (squares), hollow model (circles), concentrated model
with $A=0.5$ (up triangles), and concentrated model with $A=0.3$
(down triangles). The networks are scale-free with $\lambda=3.0$.
The data are log-binned.} \label{Ck}
\end{figure}

Equation \ref{qcl} shows that the most influential cycles on
percolation threshold are triangles; further more, for the same
number of edges forming cycles, there will be more triangles than
higher order cycles. Thus the decrease of the number of triangles
(clustering coefficients) in the same kind of networks will be
always accompanied with the drop of percolation thresholds. Figure
\ref{Cd} displays the clustering coefficient $C$\ for both normal
and hollow networks. It can be seen that the hollow networks have
much smaller clustering coefficients. Since the percolation
threshold also depend on the spectrum of the clustering
coefficient (Eq. (\ref{qc})), we examined the clustering
coefficient spectrum for the normal, hollow and concentrated
scale-free networks, and the results are shown in Fig. \ref{Ck}.
For normal networks, the spectra for $A>2$ are almost the same as
that for $A=2$. The spectra for concentrated networks follow the
same scaling law as those for normal networks, and the small
fluctuations explain the behavior of clustering coefficient in
Fig. \ref{con}(a). For hollow networks, the local clustering
coefficient is much smaller for nodes with small degrees, while it
follows the same scaling relation as that for normal networks when
degree is large. Thus the influence of the spectrum of clustering
coefficients is not crucial. Therefore the decrease in $C$ for
hollow networks is consistent with the drop of the percolation
thresholds (Figs \ref{qcle} and \ref{qc23}).

In short, we have studied how the geographical structure affects
the percolation behavior of complex networks, and provided
analytical understandings by generating function formalism on
networks with cycles. Our study gives a general suggestion on
constructing more robust real functional networks, such as the
Internet, the power grid network, etc., that arrange the edges to
connect neighboring nodes as far as possible. Although it may cost
a little more, it will stand a much reduced risk in case of node
failures. Also, the hollowing strategy could be useful to maintain
the global functions of real world networks during some
emergencies, such as epidemic occurrences, eruptions of
electronical virus, or cascade failures of power stations, etc.

L. Y. thanks the 100 Person Project of the Chinese Academy of
Sciences for their support, the Hong Kong Research Grants Council
(RGC), the Hong Kong Baptist University Faculty Research Grant
(FRG). and K.Y. thanks the Institute of Geology and Geophysics,
CAS for their support. The work is supported by the China National
Natural Sciences Foundation with Grant No. 49894190 of a major
project and the Chinese Academy of Science with Grant No.
KZCX1-sw-18 of a major project of knowledge innovation
engineering.

\end{document}